\documentstyle[12pt, aasms4, flushrt, epsfig, rotating]{article}
\lefthead{Gim et al.}
\righthead{The Open Cluster NGC 7789}

\received{}

\begin{document}

\title{The Open Cluster NGC 7789: I. Radial Velocities for Giant Stars}
\author{MUNHWAN GIM,}
\affil{\small{University of Victoria, Department of Physics and Astronomy, \\
 Victoria, BC, V8W 3P6}
\begin{center}
Electronic mail: \bf{gim@uvastro.phys.uvic.ca}
\end{center}}

\author{JAMES E. HESSER, ROBERT D. McCLURE, and PETER B. STETSON}
\affil{\small{National Research Council, 
 Herzberg Institute of Astrophysics, 
 Dominion Astrophysical Observatory  
 5071 West Saanich Road, Victoria, BC, V8X 4M6}
\begin{center}
Electronic mail: \bf{firstname.lastname@hia.nrc.ca}
\end{center}}

\begin{abstract}

A total of 597 radial-velocity observations for 112 stars 
in the $\sim$1.6~Gyr old open cluster NGC~7789 have been obtained 
since 1979 with the radial velocity spectrometer at the Dominion 
Astrophysical Observatory. 
The mean cluster radial velocity is $-$54.9 $\pm$ 0.12~km~s$^{-1}$ and 
the dispersion is 0.86~km~s$^{-1}$, from 50 constant-velocity stars selected 
as members from this radial-velocity study and the proper motion study 
of McNamara and Solomon (1981). 
Twenty-five stars (32\%) among 78 members are possible radial-velocity 
variable stars, but no orbits are determined because of the sparse sampling. 
Seventeen stars are radial-velocity non-members, while membership 
estimates of six stars are uncertain. 

There is a hint that the observed velocity dispersion falls off 
at large radius.  
This may due to the inclusion of long-period binaries preferentially  
in the central area of the cluster.
The known radial-velocity variables also seem to be more concentrated 
toward the center than members with constant velocity.
Although this is significant at only the 85\% level, when combined
with similar result of Raboud \& Mermilliod (1994) for three other
clusters, the data strongly support the conclusion that mass
segregation is being detected.
 
\end{abstract}


\section{INTRODUCTION} \label{intro}

Since stellar radial-velocity accuracies of better than 1~km~s$^{-1}$
have been attained at several observatories (e.g. \cite{GG}; \cite{L}; \cite{M}; 
\cite{Mc}), studies have been undertaken to determine membership and binary frequency 
in open clusters, as well as to estimate cluster velocity dispersions. 
With the information on both the membership and the binary stars in those clusters, 
constraints on stellar evolution, such as the effect of overshooting, 
can be improved. From the extensive observations of the red giants in open clusters 
of various ages, typical spectroscopic-binary frequencies in open clusters range 
from 25\% to 50\% (Mermilliod \& Mayor 1989, 1990; Mermilliod et al. 1995, 1996,
1997). Abt \& Willmarth (1996) note that the frequency of binaries 
changes in open clusters whose ages range from 1 $\times$ 10$^6$ to 
5 $\times$ 10$^8$ year, in the sense of increasing with age.

The velocity dispersion of the cluster giants is usually $\leq$~1~km~s$^{-1}$, 
which is consistent with simple virial-theorem estimates, so that
it is difficult to study the dynamics of open clusters based on
observations with a precision $\sim$1~km~s$^{-1}$.

The rich open cluster NGC~7789 [$\alpha$(1950)~=~23$^{h}$54\fm5, 
$\delta$(1950)~=~56\arcdeg~27\arcmin] is located at $l$~=~115\fdg49, 
$b$~=~$-$5\fdg36, and has an age $\sim$1.6~Gyr.
Its color-magnitude diagram (CMD) shows a well-defined red-giant
branch, and a populous red-giant clump, along with a large population of blue
straggler stars. The individual giant stars have been used to derive
the basic parameters such as reddening, metallicity, effective temperature,
surface gravity or masses through various photometric and spectroscopic
studies: $UBViyz$ photometry (\cite{JH}); DDO and $UBV$ 
photometry (\cite{Ja}; \cite{Cl}), $\ubvr$ photometry (\cite{Co}),
Washington photometry (\cite{CGHO}) and also moderate- and
high-resolution spectroscopy (\cite{P}; \cite{SP}; \cite{TFF}). 
An extended proper-motion membership analysis of NGC 7789 (\cite{MS}) 
identified 679 probable member stars down to B $\simeq$ 15.5 mag 
(M$_{v}$ $\approx$ 2.1) and found the existence of an extensive halo 
composed of low mass stars. The membership of some giant stars was addressed 
by the radial-velocity studies of Thogerson et al. (1993) and Scott et al. (1995).

In the present paper, for clump stars and more luminous giants in NGC~7789, 
we provide radial velocities with which we can identify membership 
and determine the spectroscopic binary frequency. 
However, the data are not extensive enough to obtain orbital elements 
for the proposed binary stars. 
Some new red giant candidates identified from CCD photometry by Gim (1998) 
were also measured to assess their membership based on their radial velocities.

The observations and data are presented in section~\ref{obs}, and analysed
in section~\ref{analy}, and the results are summarised in
section~\ref{discuss}.

\section{OBSERVATIONS and DATA} \label{obs}

Radial-velocity observations for NGC~7789 were initiated by
McClure and Hesser in 1979 with the cross-correlation radial-velocity spectrometer
(RVS) at the coud\'{e} spectrograph of the Dominion
Astrophysical Observatory (DAO) 1.2-m telescope. A detailed description 
of the DAO RVS is given by Fletcher et al. (1982) and McClure et al. (1985). 
We summarise here some points important to this project.

The RVS, following the principle of the Cambridge photoelectric
radial-velocity spectrometer (\cite{Gr}), includes a transmissive 
spectral mask, a Fabry lens and a photomultiplier. Unique designs for the Fabry 
system and spectrum mask make it possible to observe radial velocities of stars 
down to $\sim$16.0 B magnitude at the DAO 1.2-m telescope (\cite{Mc}).

The K star mask used contains 454 stellar and 14 comparison lines, covering
the wavelength range from 4332\AA~to~4765\AA~and was based on the spectrum of
Arcturus. Two image slicers were used for the observations: IS32B and
ISRVS. The former has smaller guiding errors at times of good seeing but
admits less light.

Comparison observations of a Cd-Ar source were obtained about every 30~-~60 minutes 
during each night in order to correct the zero point drift, which was usually less
than 1~km~s$^{-1}$ except for a few nights having a drift of unknown
origin up to 3~km~s$^{-1}$. 
Typically, about 10~radial-velocity standards as well as the twilight sky were
observed every night to tie our observations to the IAU system. Cluster star 
exposure times were from $\sim$15 minutes to $\sim$30 minutes, depending 
on the faintness of the stars, seeing and transparency. Scan widths were mostly 
$\sim$18~km~s$^{-1}$ for cluster stars.
Normally, this scan range covers over two-thirds of the cross-correlation dip, and
the radial velocity is calculated using a parabolic function fit to 
the dip.

The RVS data reduction procedure after observing is straightforward. The comparison
velocity is plotted against the universal time (UT) and is subtracted from the stellar
velocities for the time of observation. Secondly, a correction is made to the standard 
velocities of the DAO RVS system using radial-velocity standard values
compiled by McClure. These provide improved individual velocities, and
agree in the mean with the standard values published by Fletcher et al. (1982). 
For 30 stars in common the velocity difference (McClure $-$ Fletcher et al.) 
is $-$0.013~$\pm$~0.38~(s.d.)~km~s$^{-1}$. 

From the precision of the measurements, the internal error of the
observations is better than 0.4~km~s$^{-1}$ for stars brighter than
$B$~=~9.0 magnitude. Fletcher et al. (1982) and McClure et al. (1985)
addressed the fact that the size of the slit entrance and the seeing
are two factors affecting random error.

Overall, 597 observations for 112 stars were obtained on 71 nights from 1979 to
1996 (see Table~\ref{rvsY}). The Julian date, radial velocity and internal error 
for each star are shown in Table~\ref{rvsT} (for star IDs, see Table~\ref{rvsS}
below). For the analysis below, we have omitted seventeen observations with errors 
$\geq$ 1.0~km~s$^{-1}$ and three other observations which seem to be dubious.

The radial-velocity data are summarised in Table~\ref{rvsS}, as follows: columns 1-3,
star idenfications: K\"{u}stner (1923), McNamara \& Solomon (1981) and Gim (1998); 
columns 4-5, V and V~$-$~I photometric data from Gim's (1998) CCD photometry; 
columns 6-11, the weighted mean velocity, the standard deviation of the mean, 
number of days between first and last observations, the number of observations
used in the statistical analysis, the ratio E/I of the external over internal
(expected) errors and the probability P($\chi^{2}$) (see below); 
columns 12-14, the membership probability by proper motion (\cite{MS}) and 
by radial velocity (\cite{SFJ}), and our own membership assessment 
from this study. A detailed discussion of the membership assessment appears in the
next section.   

It is also worthwhile to note several points about the errors and the radial velocity
statistics. The errors in Table~\ref{rvsT} based on photon statistics will 
underestimate the standard deviation of the mean. Therefore the value of 
a typical RVS observational error for stars brighter than 9.0 magnitude where photon 
statistics are negligible, 0.4~km~s$^{-1}$, is added in quadrature to each error 
in Table~\ref{rvsT} before doing the statistical analysis. 
All weights are $1/error^2$. 
The P($\chi^2$) is the probability that the velocity variations are due to random
observational errors. This is estimated by assuming that the errors have 
a Gaussian distribution and are purely statistical.  
A low P($\chi^2$) value suggests the star is affected by another source of velocity 
variation, and most probably it is spectroscopic binary.

\section{RESULTS and DISCUSSION} \label{analy}

Member stars are subdivided into three groups in this study: members with constant
velocity (MC), members with variable radial velocity (MV), and members with one 
or no radial velocity data measurements in the present study (M). 
Fifty-three stars are classed MC based on P($\chi^2$)~$\geq$~0.01 
and the proper motion membership probability P($\mu$) $>$ 80\%. 
Twenty-five stars are classified as MV because they have P($\chi^2$)~$<$~0.01 
(which means that their errors come from a non-random source with 99\% confidence) 
and  because they have P($\mu$) $>$ 80\%. Although K977 has
P($\mu$)~=~0\%, it is classified  
MV because of its location on the CMD giant branch, and its mean velocity in the
present study, which is consistent with that of the cluster. 
Eleven stars with only one observation are also taken 
as members (M). All these stars have P($\mu$)~$>$~70\% and their velocities 
are within 4.7$\sigma$ of the mean cluster velocity. 
Three of them are considered radial velocity members by Scott et al. (1995).

Seventeen stars are non-members (NM) since they have 
P($\chi^2$)~$>$~0.01 and P($\mu$)~$<$~50\%. 
These stars are all lie over 7.1$\sigma$ away from the mean cluster
One NM star, K1149, has variable radial velocity. 
Four of our NM stars were also considered non-members by
Scott et al., while two others were classified as members on the basis
of radial-velocities in that study.

Membership estimates for six stars are classified
as uncertain (U). M1244 is a non-member star by P($\mu$) = 2\%, 
but its velocity from the present study is consistent with the cluster
mean. Five other stars (K193, K605, K859, M125 and M748)
have velocities from the present study that are more than 6.6$\sigma$ away 
from the mean cluster velocity, but they are considered proper
motion members by McNamara \& Solomon (1981) or radial velocity members 
by Scott et al. (1995). 

\subsection{The mean cluster velocity and its dispersion} \label{VrD}

\begin{figure}
\plotone{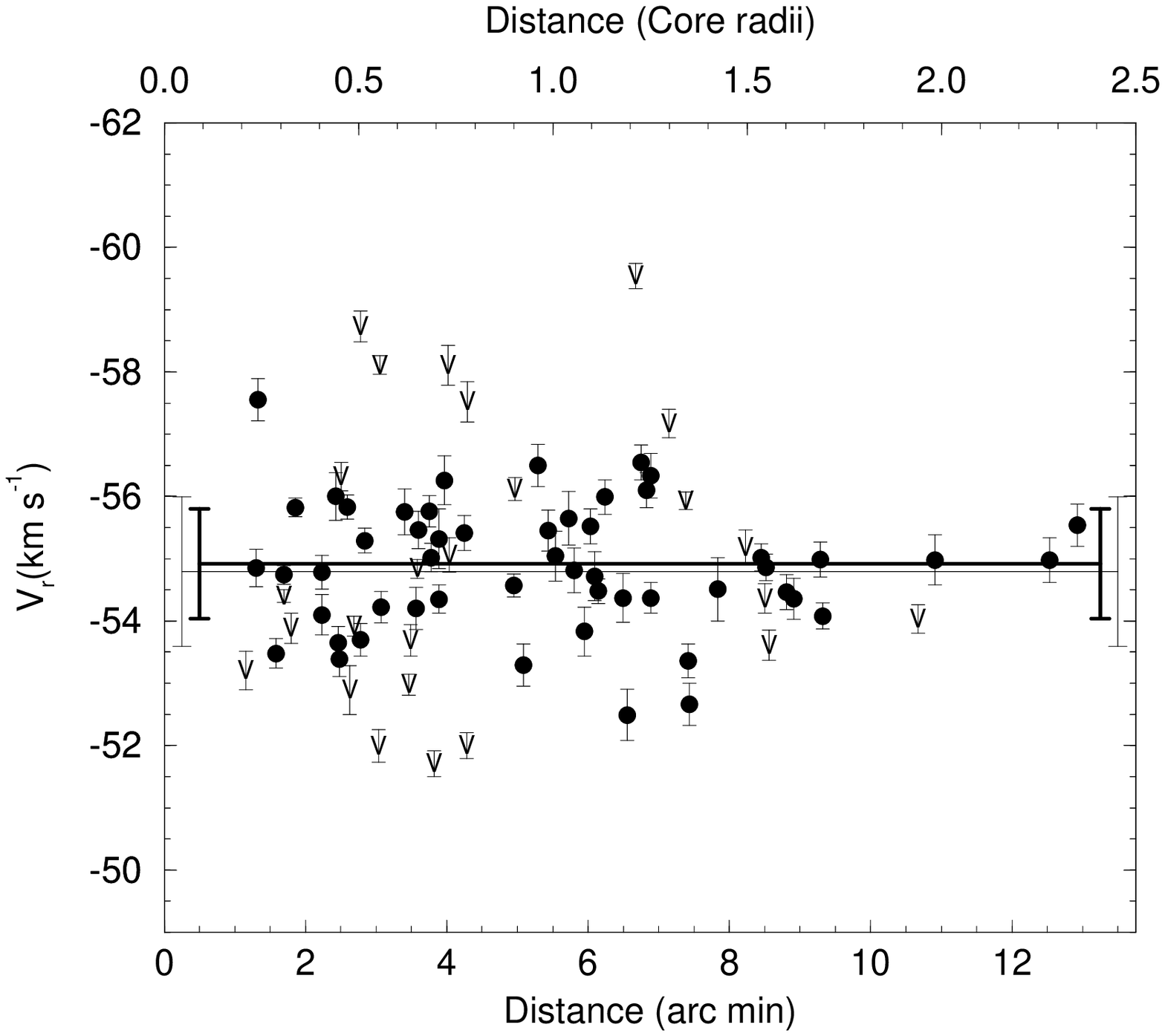}
\caption[]
{Mean radial velocity as a function of the angular distance from the
center of the cluster. 
Note that the core radius of NGC~7789 is 5\farcm5.
Solid circles:  53 members with constant radial velocities, V symbols:  
25 members with variable radial velocities.
The standard deviation of the individual velocities is shown for each 
star. The thick solid line indicates the mean cluster 
velocity, $-$54.92~$\pm$~0.12~km~s$^{-1}$, with the standard deviation 
shown at each end, $\pm$ 0.86 km s$^{-1}$. 
This is calculated from only stars with constant
radial velocities. The thin line shows the same values determined for members 
with both constant 
and variable velocities: $-$54.79~$\pm$~0.14 km~s$^{-1}$ and 1.20~km~s$^{-1}$.
}
\label{RVr}
\end{figure}

The mean cluster velocity and its dispersion were calculated for stars after excluding 
stars whose velocity is deviant by more than 2$\sigma$. 
Stars excluded are constant velocity members (MC) K526, K692 and K1114, and variable
velocity members (MV) K160, K489, K491, K865 and K897. The mean cluster velocity is 
$-$54.92~km~s$^{-1}$ and its dispersion is 0.86~km~s$^{-1}$ based on 50 MC stars.
The dispersion increases to 1.20 km s$^{-1}$ when 20 MV stars are included, while 
the mean velocity changes very little to $-$54.77~km~s$^{-1}$. We adopted, therefore,
a mean cluster velocity for NGC~7789 of $-$54.9~$\pm$~0.12~km~s$^{-1}$ and a dispersion
of 0.86~km~s$^{-1}$. Fig.~\ref{RVr} shows the radial velocities as a function of 
the distance from the cluster center for the 53 MC members and 25 MV members, 
with the adopted mean cluster velocity and the dispersion indicated by
a heavy line and error bars. 
Inspection of Fig.~\ref{RVr} suggests that outside a radius of 8 arcmin the
radial velocity dispersion is considerably lower than it is interior
to that radius. If this result is correct, it is possibly due to the presence
of binaries of very long period preferentially in the central area of the cluster. 

Below we summarise previous determinations of the mean cluster velocity.

\begin{itemize}

\item Stryker and Hrivnak (1984) found that the mean radial velocity of NGC~7789
is $-$54~km~s$^{-1}$ from three giants, K319, K415, K489, during a study of the radial
velocity variations in the  blue stragglers from image-tube spectra at a dispersion of
30~\AA~mm$^{-1}$. K319 was noticed to be a velocity variable, which is
confirmed by our observations, while we demonstrate that K489 is also
a velocity variable. 

\item Moderate-resolution spectroscopic observations (\cite{FLJ}), 
with an accuracy of 10~km~s$^{-1}$, for 12~NGC~7789 giants gave 
a mean cluster velocity of $-$57~$\pm$~7~km~s$^{-1}$. 
Our study indicates that four of their stars (K491, K637, K737, K977) 
are probable velocity variables. It is interesting that their result
and ours are similar for K605 and K859, for which we don't give any
membership assessment since the velocity of K605 is too high and our 
observational errors for K859 are too large. 

\item Recently, Scott et al. (1995) determined a mean cluster velocity of 
$-$64~km~s$^{-1}$ for NGC~7789 from spectroscopic observations of 49 member giants
with about $\pm$10~km~s$^{-1}$ accuracy. However, they adopted 
$-$57~km~s$^{-1}$ for their kinematic study in open clusters after considering 
their low accuracy and comparing with the results of Friel et al. 
(1989) and with the unpublished data of Hesser and McClure, 
which were the preliminary results of the present paper.

\end{itemize}

\subsection{The radial velocity variables}

The sparse data sampling for most of the stars, even with an observation baseline 
extending up to 6000 days, makes it difficult to determine orbital elements. 
With a 99\% confidence level that errors are non-random, 25 stars are identified 
as probable radial-velocity variable stars. The ratio of external and internal 
errors, E/I, of 20 of those stars is more than 2, considered as the minimum value 
for variability in past studies (e.g., \cite{SH}). The remaining five
have E/I $>$ 1.6.
Assuming all MV stars are binaries, the overall binary frequency is 32\% (25/78), 
which is similar to the mean binary frequency, 25\% to 38\%, 
from previous studies of open cluster, as mentioned in section~\ref{intro}.
 
Several spectroscopic studies of globular-cluster giants have shown that more velocity 
variability is found among stars near the tip of red giant branch than for stars
with fainter magnitudes. This is believed to arise from  
convective or pulsational motion in their atmosphere (\cite{PLH}; \cite{Co2}). 
However, such a trend is not found in our data for NGC~7789. 

\subsection{The apparent distribution of red giants of radial velocity variables}

\begin{figure}
\plotone{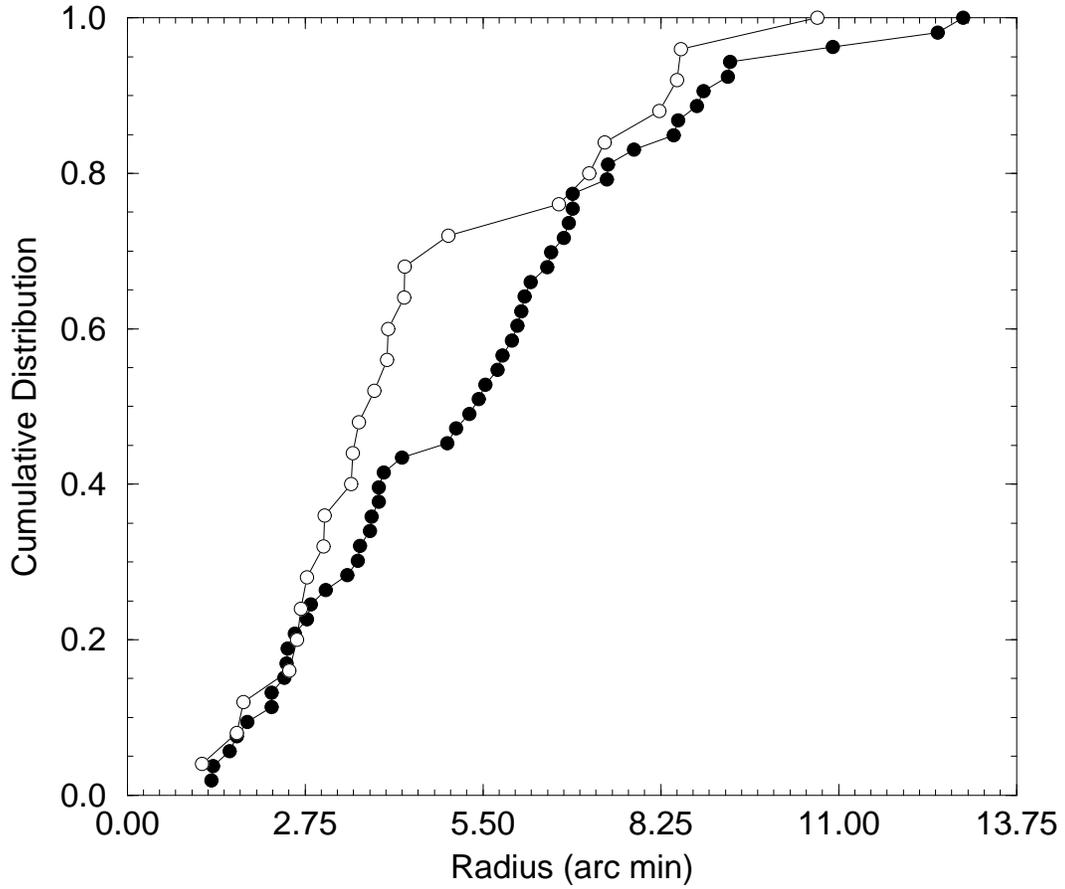}
\caption[]
{The cumulative radial distribution of stars with constant radial velocity 
(solid circles, 53 stars) and with variable radial velocity 
(open circles, 23 stars) in NGC 7789. 
The two distributions show an apparent central concentration of radial
velocity variables relative to that of stars with constant radial
velocity (see text).   
}
\label{Cum}
\end{figure}

Mass segregation during the dynamical evolution of clusters has been predicted 
by numerical simulations (\cite{SM}), and confirmed by several
studies of the radial distribution of stars with different masses (\cite{L};
\cite{A}). The cumulative distribution of stars in NGC 7789 with constant velocity 
and with variable velocity  is shown in Fig.~\ref{Cum}. Although the variable stars
appear to be more centrally concentrated, as expected if they are binaries of higher
total masses than single stars, a Kolmogorov-Smirnov test gives a
probability of only 85\%  
that the distributions of stars with constant radial velocity
(presumably single stars) 
and with variable radial velocity (presumably binaries) are
different. Note, however, that with an age $log~t$ = 9.13 (\cite{CC}),
NGC~7789 would fall in the second group of three open clusters
considered in a similar analysis by Raboud \& Mermilliod (1994). They
found the difference to be significant at the 87.9\% level. Taken
together, the joint probability that the apparent radial segregation
of spectroscopic binaries is due to chance is of order 
15\%~$\times$~12\%~$\approx$~2\%. Therefore, the combined data for
NGC~7789 (present study) and NGC~2362, 2477, 6940 (\cite{RM}) show
mass segregation at the 98\% confidence level.

\subsection{Color-magnitude diagram of the red giants in NGC 7789}

\begin{figure}
\plotone{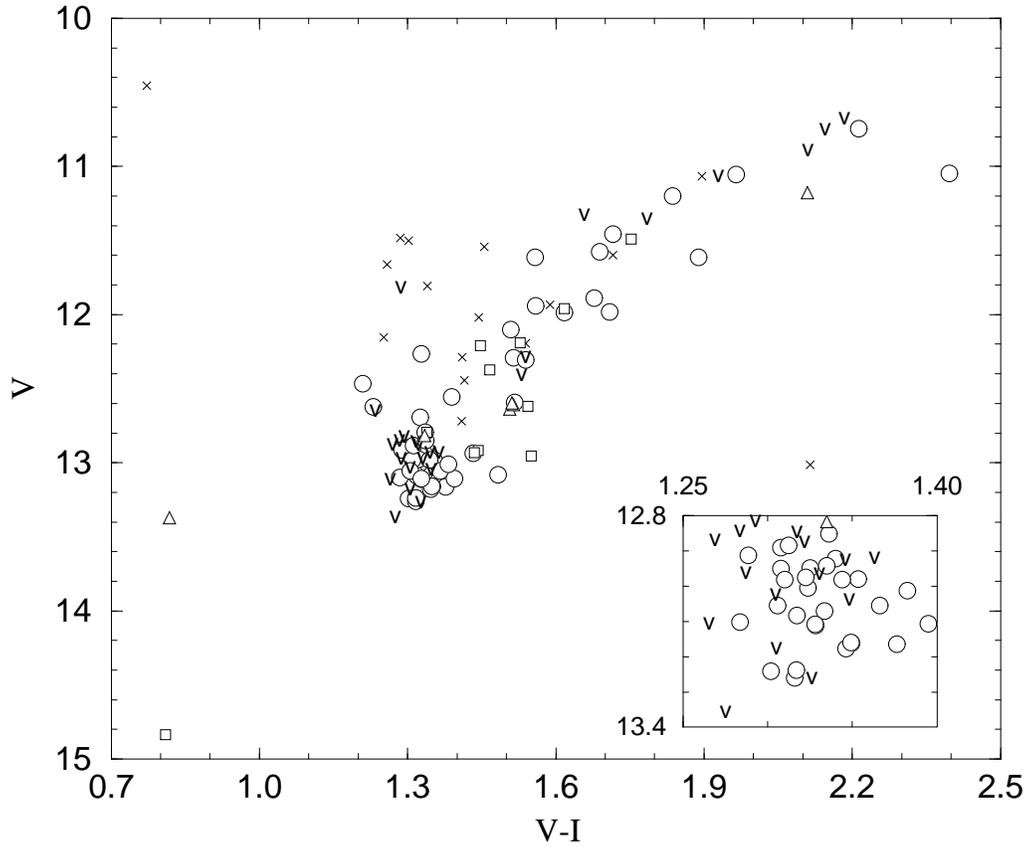}
\caption[]
{Color-magnitude diagram of the red giants in NGC 7789. 
Open circles: 53 red giant members with constant radial velocity (MC),
V marks: 25 radial-velocity variable members (MV), 
squares: 11 members with one radial velocity data (M), 
open triangles: 6 stars with uncertain membership (U),
crosses: 17 non-members (NM).
The clump star region is shown on the inset box with the expanded scale.
}
\label{CMDgiants}
\end{figure}

In the CMD of the red giants in NGC 7789 (Fig.~\ref{CMDgiants}), 
the distribution of stars in the well-defined clump and upper giant branch 
is shown with different symbols depending on their membership. 
To compare the ratio of radial-velocity variables for different populations, 
the clump stars are considered to have colors 
1.25~$<$~(V$-$I)~$<$~1.4 and magnitude 12.8~$<$~V~$<$~13.4,
while upper giant-branch stars are considered to be all stars 
which have (V$-$I)~$>$~1.2 and V~$<$~13.4 excluding the clump stars.
The frequency of radial-velocity variables for the member clump stars and 
the upper giants is almost identical (29\% for the clump stars and
28\% among the upper giants).

\section{CONCLUSION} \label{discuss}

Radial-velocity observations for 112 giant stars in NGC~7789 observed since 1979 
at the Dominion Astrophysical Observatory were used to determine cluster membership
and their radial-velocity duplicity was determined based on 
statistical analysis. Seventy-eight cluster members were selected on the basis of
radial velocities and proper motions (\cite{MS}). Among these, 25
stars (32\%) are considered to be radial-velocity variables. 
The cluster mean velocity was found to be
$-$54.9~$\pm$~0.12~km~s$^{-1}$ and its dispersion 0.86~km~s$^{-1}$ 
using 47 stars with constant velocity. Twelve additional stars with
one RVS observation were also selected as members on the basis of both 
their radial velocities and their proper motions.
Overall a frequency of 32\% for radial-velocity variables is consistent with that 
published for other clusters. The binary frequency for the more
luminous giants and clump stars is similar. The radial-velocity
variables seem to be slightly more concentrated toward the center than
stars with constant velocity. When taken alone, this result is
significant only at the 85\% level. However, three additional clusters
of the same approximate age studied by Raboud \& Mermilliod (1994) 
show the same effect, which strongly suggests 
that mass segregation is being
detected in these four clusters. More observations are needed to obtain orbital
elements for the stars which have variable radial velocities.

\acknowledgements
We wish to acknowledge NSERC for a grant to P.B.S. and J.E.H. that provided
support for M.G., and J.M. Fletcher, L. Saddlemyer and D. Bond for their
assistance over many years with the RVS. We also appreciate the helpful advice 
of J.-C. Mermilliod, S. Udry, and S. Yi in this project. M.G. is very grateful to
D.A. VandenBerg for his advice during the course of his M.Sc. program.

\begin{table}
\dummytable\label{rvsY}
\end{table}

\begin{table}
\dummytable\label{rvsT}
\end{table}

\begin{table}
\dummytable\label{rvsS}
\end{table}

\clearpage

\begin{thebibliography}{}

\bibitem[Abt 1980]{A}Abt, H.A. 1980, \apj, 241, 275

\bibitem[Abt \& Willmarth 1996]{AW}Abt, H.A., \& Willmarth, D.W. 1996, 
in The Origins, Evolution, and Destinies of Binary Stars in Clusters, ed. E.F. Milone 
and J.-C. Mermilliod, ASP Conf. Ser., 90, p. 105

\bibitem[Canterna et al. 1986]{CGHO}Canterna, R., Geisler, D., Harris, H.C., 
Olszewski, E., \& Schommer, R. 1986, \aj, 92, 79

\bibitem[Carraro \& Chiosi 1994]{CC}Carraro, G., \& Chiosi, C. 1994, \aap, 287, 761

\bibitem[Clari\'{a} 1979]{Cl}Clari\'{a}, J.J. 1979, \apss, 66, 201

\bibitem[Coleman 1982]{Co}Coleman, L.A. 1982, \aj, 87, 369

\bibitem[C\^{o}t\'{e} et al. 1996]{Co2}C\^{o}t\'{e}, P., Pryor, C., McClure, R.D., 
Fletcher, J.M., \& Hesser, J.E. 1996, \aj, 112 574

\bibitem[Fletcher et al. 1982]{F}Fletcher, J.M. Harris, H.C., McClure, R.D., 
\& Scarfe, C.D. 1982, \pasp, 94, 1017 

\bibitem[Friel et al. 1989]{FLJ}Friel, E.D., Liu, T., \& Janes, K.A. 1989, \pasp, 
101, 1105

\bibitem[Gim 1998]{G}Gim, M. 1998, M.Sc. Thesis, University of Victoria  

\bibitem[Griffin 1967]{Gr}Griffin, R.F. 1967, \apj, 148, 465

\bibitem[Griffin \& Gunn 1974]{GG}Griffin, R.F., \& Gunn, J.E. 1974, \apj, 191, 545

\bibitem[Janes 1977]{Ja}Janes, K.A. 1977, \aj, 82, 35

\bibitem[Janes \& Phelps 1994]{JP}Janes, K.A., \& Phelps, R.L. 1994, \aj, 108, 1773

\bibitem[Jennens \& Helfer 1975]{JH}Jennens, P.A., \& Helfer, H.L. 1975, \mnras, 
172, 681

\bibitem[K\"{u}stner 1923]{K}K\"{u}stner, F. 1923, Bonner Ver\"{o}ff, No. 18

\bibitem[Latham 1985]{L}Latham, D.W. 1985, in Stellar Radial Velocities, ed. A.G.D. 
Philip and D.W. Latham (Schenectady, L. Davis), p. 21

\bibitem[Mayor 1985]{M}Mayor, M. 1985, in Stellar Radial Velocities, ed. A.G.D. 
Philip and D.W. Latham (Schenectady, L. Davis), p. 35

\bibitem[McClure et al. 1985]{Mc}McClure, R.D., Fletcher, J.M., 
Grundman, W.A., \& Richardson, E.H. 1985, in Stellar Radial Velocities, ed. A.G.D. 
Philip and D.W. Latham (Schenectady, L. Davis), p. 49 

\bibitem[McNamara \& Solomon 1981]{MS}McNamara, B.J., \& Solomon, S. 1981, 
\aaps, 43, 337 

\bibitem[Mermilliod \& Mayor 1989]{MM1}Mermilliod, J.-C., \& Mayor, M. 1989, 
\aap, 219, 125

\bibitem[Mermilliod \& Mayor 1990]{MM2}Mermilliod, J.-C., \& Mayor, M. 1990, 
\aap, 237, 61

\bibitem[Mermilliod et al. 1995]{MAN}Mermilliod, J.-C., Anderson, J., 
Nordstr\"{o}m, B., \& Mayor, M. 1995, \aap, 299, 53

\bibitem[Mermilliod et al. 1996]{MHRM}Mermilliod, J.-C., Huestamendia, G., 
del Rio, G., \& Mayor, M. 1996, \aap, 307, 80

\bibitem[Mermilliod et al. 1997b]{MCAM}Mermilliod, J.-C., Clari\'{a}, J.J., 
Anderson, J., \& Mayor, M. 1997, \aap, 324, 91


\bibitem[Pilachowski 1985]{P}Pilachowski, C.A. 1985, \pasp, 97, 801

\bibitem[Pryor et al. 1988]{PLH}Pryor, C.P., Latham, D.W., \& Hazen, M.L. 
1988, \aj, 96, 123

\bibitem[Raboud \& Mermilliod 1994]{RM}Raboud, D., \& Mermilliod, J.-C. 1994, 
\aap, 289, 121

\bibitem[Scott et al. 1995]{SFJ}Scott, J.E., Friel, E.D., \& Janes, K.A. 1995, 
\aj, 109, 1706

\bibitem[Sneden \& Pilachowski 1986]{SP}Sneden, C., \& Pilachowski, C.A. 1986, 
\apj, 301, 860

\bibitem[Spitzer \& Mathieu 1980]{SM}Spitzer, L., Jr., \& Mathieu, R.D. 1980, 
\apj, 241, 618

\bibitem[Stryker \& Hrivnak 1984]{SH}Stryker, L.L., \& Hrivnak, B.J. 1984,
\apj, 278, 215

\bibitem[Thogersen et al. 1993]{TFF}Thogersen, E.N., Friel, E.D., 
\& Fallon, B.V. 1993, \pasp, 105, 1253
\end{thebibliography}
\end{document}